**Correspondence to:**
R. E. Hibbins,
robert.hibbins@ntnu.no

# SuperDARN Observations of Semidiurnal Tidal Variability in the MLT and the Response to Sudden Stratospheric Warming Events

R. E. Hibbins[1,2], P. J. Espy[1,2], Y. J. Orsolini[2,3], V. Limpasuvan[4], and R. J. Barnes[5]

[1]Department of Physics, Norwegian University of Science and Technology (NTNU), Trondheim, Norway, [2]Birkeland Centre for Space Science, Bergen, Norway, [3]Norwegian Institute for Air Research (NILU), Kjeller, Norway, [4]Coastal and Marine Systems Science, Coastal Carolina University, Conway, SC, USA, [5]Applied Physics Laboratory, The Johns Hopkins University, Baltimore, MD, USA

**Abstract** Using meteor wind data from the Super Dual Auroral Radar Network (SuperDARN) in the Northern Hemisphere, we (1) demonstrate that the migrating (Sun-synchronous) tides can be separated from the nonmigrating components in the mesosphere and lower thermosphere (MLT) region and (2) use this to determine the response of the different components of the semidiurnal tide (SDT) to sudden stratospheric warming (SSW) conditions. The radars span a limited range of latitudes around 60°N and are located over nearly 180° of longitude. The migrating tide is extracted from the nonmigrating components observed in the meridional wind recorded from meteor ablation drift velocities around 95-km altitude, and a 20-year climatology of the different components is presented. The well-documented late summer and wintertime maxima in the semidiurnal winds are shown to be due primarily to the migrating SDT, whereas during late autumn and spring the nonmigrating components are at least as strong as the migrating SDT. The robust behavior of the SDT components during SSWs is then examined by compositing 13 SSW events associated with an elevated stratopause recorded between 1995 and 2013. The migrating SDT is seen to reduce in amplitude immediately after SSW onset and then return anomalously strongly around 10–17 days after the SSW onset. We conclude that changes in the underlying wind direction play a role in modulating the tidal amplitude during the evolution of SSWs and that the enhancement in the midlatitude migrating SDT (previously reported in modeling studies) is observed in the MLT at least up to 60°N.

## 1. Introduction

Classical atmospheric tidal theory (Chapman & Lindzen, 1970) predicts that the strongest tidal mode in the midlatitude to high latitude is the first symmetric mode of the migrating (i.e., Sun-synchronous) semidiurnal tide (SDT). This tide is primarily generated in the stratosphere through the day-night cycle of solar insolation on low-latitude ozone. Conservation of energy dictates that the amplitude of the tide grows exponentially as it propagates vertically such that, in the midlatitude to high-latitude mesosphere and lower thermosphere (MLT) region, the migrating SDT is often the largest amplitude periodic oscillation observed in the horizontal wind.

Tide perturbations play a crucial role in distributing atmospheric trace constituents (e.g., Ward, 1998), in modulating gravity wave (GW) fluxes (e.g., Andrioli et al., 2013; Beldon & Mitchell, 2010; Espy et al., 2004), and in the momentum budget of the lower thermosphere (e.g., Becker, 2017; Miyahara et al., 1993). Changes in the forcing conditions in the lower atmosphere, coupling between different meridional and zonal modes of the tide, tidal interactions with planetary waves (PWs) and/or GWs, and changes in the background atmosphere (through which the tide propagates) can all modulate the observed tidal amplitude over time (Riggin et al., 2003).

It is often assumed that the semidiurnal wind variability is dominated by the migrating SDT (hereafter, SW2 for semidiurnal, westward propagating, zonal wavenumber 2) since it is theoretically predicted to be strongest beyond the tropics. However, observational studies using multiple ground-based radars have demonstrated that nonmigrating components other than the pure SW2 are often present in the Northern Hemisphere (NH) MLT region (e.g., Manson et al., 2009; Portnyagin et al., 2004). Using the NASA Thermosphere Ionosphere Mesosphere Energetics and Dynamics satellite Doppler Interferometer (TIDI) measurements, Iimura et al. (2010) demonstrated that a nonmigrating SW1 (i.e., semidiurnal, westward propagating zonal wavenumber 1 tide) is clearly present in the NH MLT horizontal winds, maximizing around 60°N in late spring/early summer. In addition, evidence was presented for the presence of a nonmigrating wavenumber 3 (SW3) and a weak zonally symmetric (S0) SDT especially evident in the lower thermosphere. Although the observed NH nonmigrating tides are generally weaker than those observed in the Southern Hemisphere MLT during the austral summer months (e.g., Baumgaertner et al., 2006; Hibbins et al., 2010; Iimura et al., 2009; Murphy et al., 2006), their contributions to the total NH MLT SDT variance are significant.

The origins of these nonmigrating tides are still under debate (Miyoshi et al., 2017). Zonally asymmetric latent heating (e.g., Hagan & Forbes, 2003) and nonlinear interactions between migrating tides and PWs may both play a part in their generation (e.g., Angelats i Coll & Forbes, 2002; Teitelbaum & Vial, 1991). However, the forcing mechanisms and the propagation characteristics of each mode are clearly very different.

Recent observational studies have noted SDT enhancements in response to changes in the NH polar upper atmosphere associated with the sudden stratospheric warming (SSW) phenomenon (e.g., Bhattacharya et al., 2004; Goncharenko et al., 2012; Nozawa et al., 2012; Orsolini et al., 2017). During SSWs, the wintertime circumpolar stratospheric wind can reverse, and the polar stratosphere rapidly warms on the order of 10K over a few days. The anomalous mean meridional circulation (MMC) induced by strong quasi-stationary PW forcing during SSWs can impose far-reaching effects vertically, both below and above the stratosphere, as well as meridionally toward the summer hemisphere (e.g., de Wit et al., 2015; Randel, 1993). Observations using a single medium-frequency radar at Tirunelveli (9°N, 78°E) by Sridharan et al. (2012) revealed enhancements in the semidiurnal wind variability in the low-latitude MLT during the 2006 and 2009 SSW events. Tidal enhancements have also been observed in the low-latitude ionosphere during the 2009 SSW (e.g., Lin et al., 2012; Pedatella & Forbes, 2010). Pedatella and Forbes (2010) suggested that the interaction between the migrating SDT and strong PW activity during SSWs promoted the amplification of nonmigrating SDT components.

Previous global climate model simulations suggest modulation of the SDT in the midlatitude to high latitude during SSW events (e.g., Sassi et al., 2013). Using the National Center for Atmospheric Research Whole Atmosphere Community Climate Model with specified dynamics (SD-WACCM; Marsh, 2011; Marsh et al., 2013), Limpasuvan et al. (2016) modeled a robust amplification of the SW2 tide in the composite MLT structure during SSWs. Focusing particularly on SSWs with an associated elevated stratopause (ES) when strong stratosphere-MLT coupling persists, these authors illustrated that the enhanced equatorial upwelling (and cooling) due the anomalous MMC led to an increase in tropical middle and upper stratospheric ozone, as observed by Sridharan et al. (2012). The region of increased ozone concentration coincides with enhanced upward SW2 wave activity that ultimately propagates into the high winter latitude after the SSW onset (when the stratospheric wind reverses at 1 hPa). In analyzing the SD-WACCM simulation for the 2013 SSW-ES and the 12-hr variability in the horizontal wind observed from meteor radar observations over Trondheim, Norway (63.4°N), Orsolini et al. (2017) also provided local observational evidence of SDT enhancement in the high-latitude MLT. The enhanced SDT structure peaked at a lower altitude range and exhibited a much larger amplitude than the corresponding SW2 simulated in SD-WACCM. However, as noted by Smith (2012), SD-WACCM tends to underestimate tidal signatures.

While the several aforementioned studies show the excitation of the SDT at high winter latitudes following SSWs, collaborative observations are required to lend further confidence to these simulations. However, these observations are hard to attain. Single-station measurements are unable to characterize the separate zonal components of the SDT and their variability. Constructive and destructive interference between different zonal wavenumbers can in itself drive longitudinal and temporal variations in the locally measured semidiurnal wind variability (e.g., Hibbins et al., 2010). Any discussion of the drivers of tidal variability needs to address each mode separately. Furthermore, the behavior of nonmigrating modes of the SDT during SSWs is still uncertain. In order to isolate the different tidal modes, and to measure their temporal evolution over short time scales, high quality longitudinally spread contemporaneous long-term observations are required, ideally from cross-calibrated systems of a similar consistent engineering design.

The NH midlatitude to high-latitude chain of Super Dual Auroral Radar Network (SuperDARN; Greenwald et al., 1995) radars, consisting of a chain of eight high-frequency (HF) radars that span nearly 180° of

Table 1
*Summary of the Eight NH SuperDARN Radars Used in This Study*

| Radar | Code | Longitude (°) | Latitude (°) | Start date | End date |
|---|---|---|---|---|---|
| Hankasalmi | HAN | +25.2 | 64.4 | 1995-02-22 | 2016-03-13 |
| Þykkvibaer | PYK | −18.0 | 65.7 | 1995-11-20 | 2016-03-13 |
| Stokkseyri | STO | −26.9 | 64.7 | 1994-08-29 | 2016-03-13 |
| Goose Bay | GBR | −60.3 | 55.5 | 1993-09-29 | 2016-03-19 |
| Kapuskasing | KAP | −83.3 | 51.4 | 1993-09-29 | 2016-03-24 |
| Saskatoon | SAS | −105.2 | 54.2 | 1993-09-29 | 2016-03-22 |
| Prince George | PGR | −123.2 | 56.1 | 2000-03-03 | 2016-03-21 |
| Kodiak | KOD | −150.1 | 59.5 | 2000-01-08 | 2013-06-11 |

*Note.* Coordinates represent the approximate geographic center of the first four range gates of each radar. Dates are formatted as YYYY-MM-DD.

longitude, can be used to delineate the various zonal components of the SDT. Kleinknecht et al. (2014) demonstrated that concurrent horizontal wind data associated with the meteor drift observed by SuperDARN radars could be used to derive the various PW components (zonal wavenumbers 1 and 2) in the MLT at these latitudes. Climatologies (Kleinknecht et al., 2014) and studies of PW variability during summer (Stray, Espy, et al., 2015), the autumn equinox (Stray et al., 2014), and during SSWs (Stray, Orsolini, et al., 2015) were presented using SuperDARN data recorded between 2000 and 2008.

Using an approach similar to that presented in Kleinknecht et al. (2014), we demonstrate in this paper that meteor wind data from the NH midlatitude to high-latitude chain of SuperDARN radars can be used to extract and isolate the primary (migrating and nonmigrating) components of the NH MLT SDT. To date, simultaneous data from multiple SuperDARN radars have not been used in this manner. Derived from over 20 years of radar data, climatologies of the different components of the SDT are presented to facilitate additional discussion on the drivers of tidal variability. Second, using a composite of 13 SSW-ES events, we present observational evidence on the role of SSWs in modulating the components of the SDT in the NH midlatitude to high-latitude wintertime when this tide is known to be both strong and highly variable.

## 2. Data and Methodology

The HF radars in the SuperDARN network measure electric fields in the ionospheric *E* and *F* regions in the midlatitude and high-latitude regions of both hemispheres (Greenwald et al., 1995). Using the multipulse sounding technique of Farley (1972), the radars also detect grainy near-range echoes in the lowest range gates that are predominately scattered from meteor trails whose velocity data correspond to the neutral wind near 95 km (Hall et al., 1997). For this study, we focus on the SuperDARN chain of eight NH radars that span a limited latitudinal range over nearly 180° of longitude, from Hankasalmi in Finland to Kodiak in Alaska (Table 1).

Figure 1 shows the data coverage from the radar chain during this period. Simultaneous data from at least six individual radars are typically available on any given day after the mid-1990s.

We adopt an approach similar to Kleinknecht et al. (2014) to extract various zonal wavenumber components of the MLT SDT from SuperDARN meteor wind data recorded between 1995 and 2016. The horizontal components of the hourly mean winds are fitted with a single value decomposition (Press et al., 1992) to the line-of-sight meteor trail velocities detected in the first four range gates of each radar (Hibbins & Jarvis, 2008). These gates lie between 180- and 400-km distance from the radar (Chisham & Freeman, 2013) and intercept the upper MLT where these HF radars typically detect meteor ablation trails. Due to each individual radar's limited spread in azimuth (~45°) and the poleward orientation of the majority of radars in the chain, the meridional wind component is typically fitted with a lower standard deviation and is therefore the component presented in the subsequent analyses. The derived meridional winds are representative of the altitude range over which meteor trails are detected at around 12 MHz (Chisham & Freeman, 2013) and agree well with horizontal winds recorded at around 95 km altitude as determined from comparative studies with other colocated radars (Hall et al., 1997; Hibbins & Jarvis, 2008; Hussey et al., 2000).

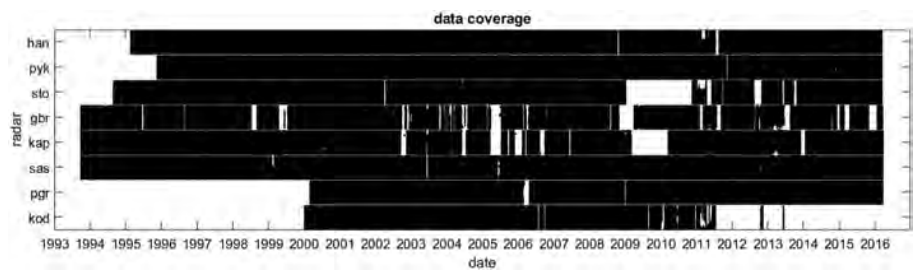

**Figure 1.** Data coverage (black) from the eight Northern Hemisphere SuperDARN radars used in this study. The bar for each radar represents hour of day (0–24 UT) up the ordinate and date along the abscissa. Three-letter radar codes are taken from Table 1 and are displayed from east to west (with the westernmost radar at the bottom).

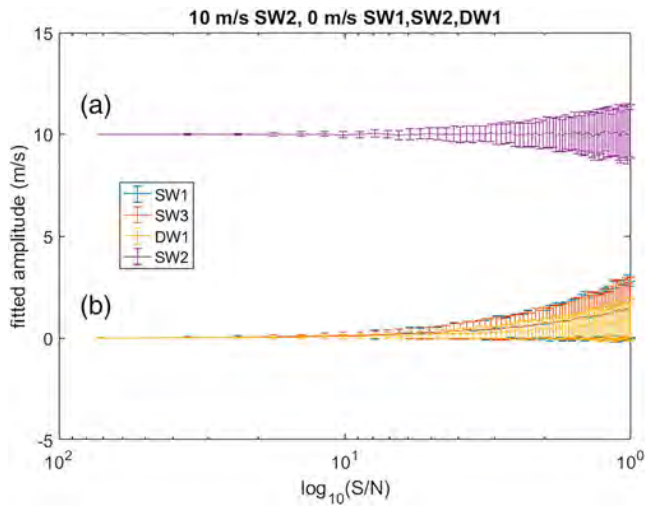

**Figure 2.** Example four-component fit to trial data of varying signal-to-noise ratios. In this case the four components used in the paper are fitted to a pure 10 m/s amplitude SW2 tide over a range of white (Gaussian) noise levels. Error bars represent twice the standard deviation of the fitted amplitudes. DW1 = migrating diurnal, zonal wavenumber 1 tide; SW1 = semidiurnal, westward propagating, zonal wavenumber 1 tide; SW2 = semidiurnal, westward propagating, zonal wavenumber 2 tide; SW3 = semidiurnal, westward propagating, zonal wavenumber 3 tide.

The hourly mean meridional winds underwent a quality control procedure similar to that discussed in Kleinknecht et al. (2014) before an average 24-hr day was generated for each radar, obtained by averaging each of the hourly mean winds derived over a running 4-day period with each data point weighted by the reciprocal of the square of the standard deviation of the hourly mean fitted wind. A total of at least 48 hr of data were required in each 4-day period, with each hour repeated at least twice. The mean wind was subtracted from each mean day for each radar to generate a residual mean wind. A nonlinear least squares surface fit was then performed on the residual winds from all radars given the hour of day and longitude of the observations, provided data from at least four radars were present in any given 4-day period. The functional form of the fit represented the migrating diurnal, zonal wavenumber 1 tide (DW1) and migrating semidiurnal, zonal wavenumber 2 tide (SW2) together with the nonmigrating, semidiurnal westward propagating zonal wavenumber 1 (SW1) and wavenumber 3 (SW3) tides. During the fitting process each residual wind data point was weighted by the reciprocal of the square of the standard deviation of the mean measurement.

In order to investigate the independence of each of the fitted components, tests were undertaken as follows: In each of four cases artificial data were generated at each radar longitude representative of a single component of the tides (DW1, SW1, SW2, and SW3) of amplitude 10 m/s. This was subjected to the four-component fitting process described above. Increasing levels of white (Gaussian) noise were added to the data to simulate a range of realistic signal-to-noise ratios (S/N) between 100 and 1. Each fitting routine was run 100 times with 100 different realizations of white noise, and the final fitted amplitudes of the four components together with the standard deviation of the fit were investigated. An example of the four-component fit generated from a pure SW2 wave is reproduced in Figure 2 over the range of S/N.

These simulations indicated that there was no significant cross-talk between each of the four fitted modes of the tides. As the data tend toward a S/N of 1 the components not present in the synthetic data grow in fitted amplitude, but in no case was their fitted amplitude significantly different from 0. And in no cases did the fitted amplitude of the true component systematically deviate from its correct value; only the error in the fitted amplitude grows in line with the higher noise level in the synthetic data. Similar results were obtained for four component fits to each of the three other individual tidal components.

Iimura et al. (2010) detected a small S0 component to the SDT in their analysis of the nonmigrating components of the tide in the TIDI satellite data. Due to the limited spread in longitudes of the radar chain, we are unable to unambiguously fit the S0 component of the SDT. Tests showed that an S0-like response can be returned from the fitting procedure as a linear combination of the SW1, SW2, and SW3 SDTs over the longitudes of the radar chain. So undersampling of a potential S0 component may oversample the other components, especially the SW1 and SW2. This must be considered in interpreting the data. However, we note that the S0 component was the smallest nonmigrating component of the SDT extracted from the TIDI meridional winds at 60°N and 95 km altitude (Iimura et al., 2010). We also see no strong evidence that the SW1 and SW2 components vary systematically, either in the climatologies presented here or in the amplitudes of the daily fits.

Figure 3 shows an example of the surface fit to the hourly mean meridional wind recorded from the eight radars (orange circles) simultaneously over the 4-day period centered on 18 September 2008. In this case, the tidal disturbance was predominantly SW2, precessing westward at a rate of 180° of longitude every 12 hr (i.e., Sun-synchronous). To examine the robustness of the fitting process, the nonlinear least squares fit was repeated eight times, each time with all data from one of the eight radars omitted from the fit. These tests showed that the overall quality of the fit and the outcome of the fitting process does not depend on data from any one particular radar. Figure 3 illustrates the advantage of using the simultaneous radar chain; unlike single-station observations or uncoordinated multiple stations, the derived tidal perturbation is global with zonal wavenumber 1–3 migrating and nonmigrating components potentially well delineated.

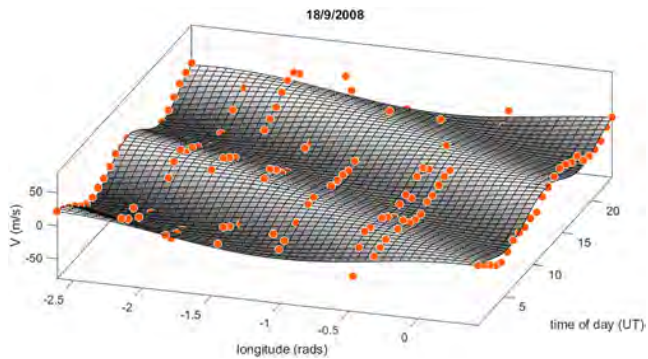

**Figure 3.** Example of a surface fit (gray plane) to 1 day of hourly mean meridional wind from the eight SuperDARN radars (orange circles). Data are plotted as a function of radar longitude and UT of day.

## 3. Climatology of the SuperDARN SDTs

Time series of the 4-day running mean amplitude of the three fitted components of the SDT (SW1, SW2, and SW3) are presented in Figures 4a and 4b. The figures clearly show that the SW2 (migrating) component of the tide is typically stronger than the other components with a repeatable annual cycle peaking in winter and autumn superimposed on a longer-term amplitude modulation with peaks around the late 1990s and the years around 2013. The red vertical lines in Figure 4 represent the onset dates of SSWs (with an associated ES) taken from Limpasuvan et al. (2016) and will be discussed in the next section.

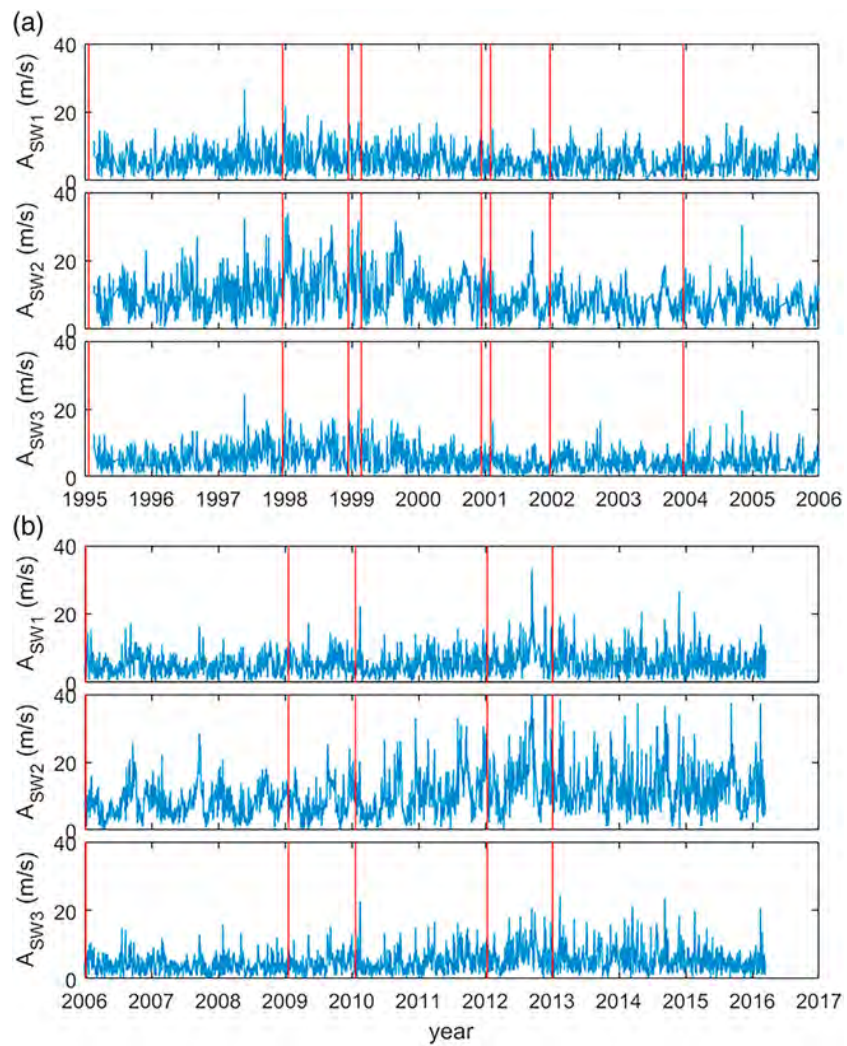

**Figure 4.** (a) Daily mean amplitude of the three components of the semidiurnal tide (blue) over the period 1995 to 2005 around 60°N (top = SW1; middle = SW2; bottom = SW3). The red vertical lines represent the onset dates of sudden stratospheric warming (with an elevated stratopause) taken from Limpasuvan et al. (2016). (b) As in (a), but covering the period 2006 to 2017. SW1 = semidiurnal, westward propagating, zonal wavenumber 1 tide; SW2 = semidiurnal, westward propagating, zonal wavenumber 2 tide; SW3 = semidiurnal, westward propagating, zonal wavenumber 3 tide.

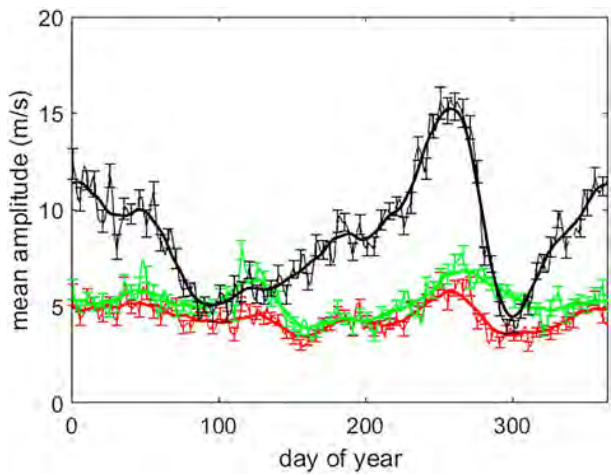

**Figure 5.** Climatology of the amplitude of the nonmigrating SW1 (green), migrating SW2 (black), and nonmigrating SW3 (red) components of the semidiurnal tide around 60°N based on all data between 1995 and 2016. The thin line with errors represents the daily mean amplitude, and the thick line is a 30-day running mean. Error bars (representing the standard error of the mean) are only shown for every fifth data point for clarity.

Climatologies of these three components are shown in Figure 5. In common with previous studies, the SDT amplitude at these latitudes in the MLT maximizes around the autumn equinox with a secondary maximum in wintertime. Mitchell et al. (2002) observed maxima in the semidiurnal wind variability of >30 m/s amplitude in the meridional wind field in the autumn and winter from single-station meteor radar observations around 95 km above Esrange (68°N, 21°E). Similar values have been reported around these altitudes and latitudes from satellite observations (e.g., Burrage et al., 1995) and other ground-based radars (e.g., Kishore Kumar & Hocking, 2010). We note that although the overall climatology is similar in form to other observations from these latitudes, the tidal amplitudes reported here are somewhat smaller. It should be noted that the SuperDARN radars derive a mean hourly wind over a vertical full width at half maximum around 12 km (Chisham & Freeman, 2013), and hence the measured amplitudes of waves with small vertical wavelengths may be underreported in the derived data from SuperDARN radars compared to other systems with height finding and higher vertical resolution (see, e.g., Hibbins et al., 2011, for comparisons of tide data derived with a single SuperDARN radar compared to that derived from a near colocated Skiymet system).

Figure 5 shows that the two maxima in autumn and winter are predominantly due to the migrating (SW2) component of the SDT with smaller contributions from SW1 and SW3 especially around the equinoxes. The SW1 and SW3 components appear to vary similarly, with the amplitude of the SW1 component slightly larger throughout most of the year. A striking feature of these resolved components is the rapid reduction in the strength of the SW2 component after the autumn equinox coincident with an increase in the strength of the SW1 component. Figure 5 demonstrates that during days 280–310, the SW1 component is actually the strongest contribution to the observed 12-hr variability at these latitudes, and during springtime (between approximately days 80 and 140) the three zonal components are of a similar amplitude. By comparing radar data from Svalbard (78°N, 18°E) and the CANDAC-PEARL station (80°N, 86°W), Manson et al. (2011) observed a weak, but significant, SW1 SDT which was strongest around March–June, together with a mix of SW1 and SW3 SDTs in late autumn/winter in addition to the dominant migrating (SW2) SDT. This is in good agreement with the results presented here from slightly lower latitudes and also with the TIDI satellite results of Iimura et al. (2010) from around 60°N discussed earlier.

Figures 4 and 5 demonstrate a complex mixture of semidiurnal tidal modes throughout the year around the NH SuperDARN radars' latitude (approximately 60°N). The investigation of the drivers of the nonmigrating tidal components and their long-term variability will be the subject of future work. In the next section, we concentrate on the role of SSWs in driving the short-term variability of the three components of the SDT in light of the modeling work of Limpasuvan et al. (2016).

## 4. SDT Variability During the 2013 SSW

We begin by comparing the response of the SDT to the 2013 SSW presented in Orsolini et al. (2017) from the Trondheim (63°N) meteor radar with the three individual components of the SDT derived from the SuperDARN data corresponding to a similar latitude. Figure 6 shows the amplitude of the 12-hr meridional wind variability presented as a contour plot between 82 and 97 km over Trondheim for a 5-week period beginning 1 week before the onset date of the 2013 SSW (5 January 2013, as defined below). For comparison, the three components of the SDT derived from the SuperDARN chain of radars are presented over exactly the same time period.

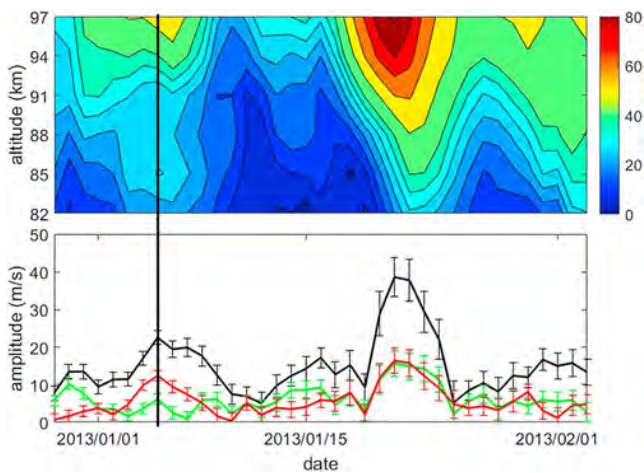

**Figure 6.** (top panel) Four-day running mean amplitude of the 12-hr variability in the meridional wind observed with the SKiYMET meteor radar at Trondheim, Norway (63°N) between 82- and 97-km altitude (data from Orsolini et al., 2017). (bottom panel) The three components of the semidiurnal tide resolved from the longitudinal chain of SuperDARN radars (green = SW1; black = SW2; red = SW3). Data span the 2013 sudden stratospheric warming over a 5-week interval starting 1 week before the onset date of 5 January 2013 which is indicated by a vertical black line.

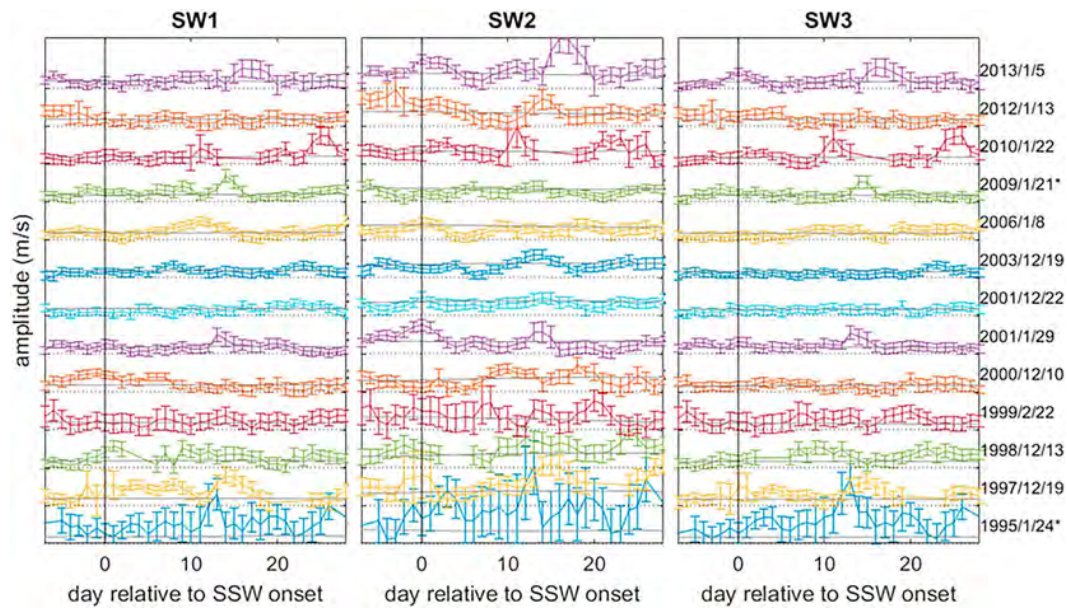

**Figure 7.** Amplitude of the three components of the semidiurnal tide (color error bar plots) over the 13 SSW events associated with an elevated stratopause identified by their onset dates (right axis, right-hand panel) plotted over 5 weeks beginning 1 week before the onset. Left panel = SW1; middle panel = SW2; right panel = SW3. The onset dates marked with an asterisk represent vortex split events. The pale gray line on each panel represents the smoothed climatology of the tidal amplitude for each event. The height of each panel, separated by dotted lines, represents 30 m/s amplitude. SSW = sudden stratospheric warming; SW1 = semidiurnal, westward propagating, zonal wavenumber 1 tide; SW2 = semidiurnal, westward propagating, zonal wavenumber 2 tide; SW3 = semidiurnal, westward propagating, zonal wavenumber 3 tide.

Consistent with the single-station meteor radar observations from Trondheim presented in Orsolini et al. (2017), a strong burst of SDT is clearly seen in the SuperDARN SDT components following the SSW peaking at around 17 days after onset. These data demonstrate that the enhancement is associated strongly with the SW2 component, as predicted by the SD-WACCM modeling presented in Orsolini et al. (2017), and to a lesser extent with the nonmigrating components.

## 5. The Mean Response of the SDT to Recent SSWs

The red vertical lines in Figure 4 represent the onset dates of major SSWs with associated ES events, as taken from Limpasuvan et al. (2016). Briefly, based on the criteria established in Stray, Orsolini, et al. (2015), an SSW event is declared when three conditions are met: (1) The wintertime polar cap (70°N to 90°N) averaged temperature drops below 190 K between 80 and 100 km; (2) the polar cap zonal-mean zonal wind reverses from eastward to westward at 1 hPa (~50 km) and persists for longer than 5 days (Tweedy et al., 2013); and (3) the polar cap stratopause altitude changes by at least 10 km upward. The 13 SSW onset dates span the period between January 1995 and January 2013. Two winters (1998/1999 and 2000/2001) experienced two SSWs with onsets separated by 70 and 49 days, respectively. Of these 13 SSWs, only two are categorized as vortex split events (onset dates 24 January 1995 and 21 January 2009) based on SD-WACCM geopotential maps at 10 hPa (Limpasuvan et al., 2016), with the others being vortex displacement events. Vortex displacement during SSWs refers to the occurrence when the vortex structure becomes highly displaced off the pole for an extended period. The geopotential height field that characterizes the vortex takes on a zonal number 1 characteristic, with one local minimum in the midlatitude to high latitude. Vortex split during SSWs refers to the rarer occurrence when the polar vortex evolves into two distinct vortices with two local minima in the geopotential height field.

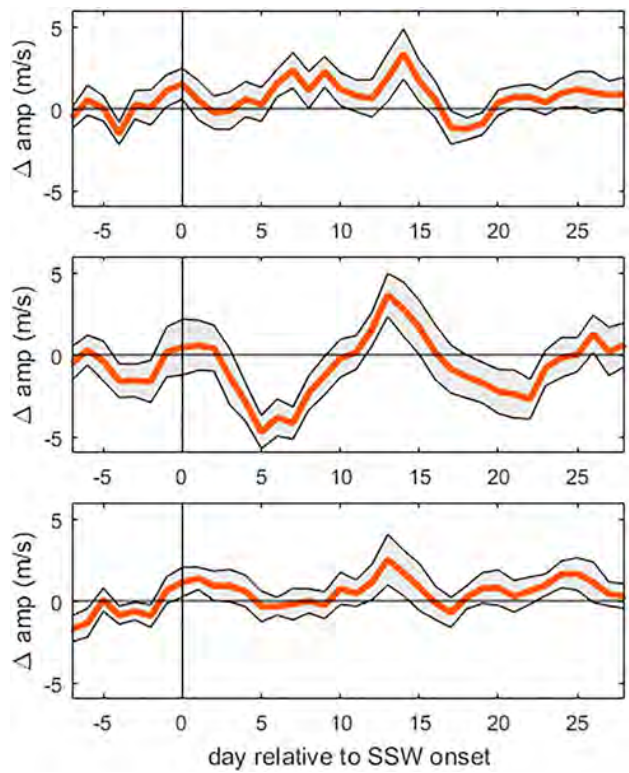

**Figure 8.** The composite behavior of the anomalous amplitude of the three semidiurnal tide components around 60°N and 95 km (top panel = SW1; middle panel = SW2; bottom panel = SW3). The composite is based on the 13 sudden stratospheric warmings presented in Figure 7. The gray area represents the standard error of the weighted mean, and the vertical black line represents the sudden stratospheric warming onset at day 0.
SW1 = semidiurnal, westward propagating, zonal wavenumber 1 tide;
SW2 = semidiurnal, westward propagating, zonal wavenumber 2 tide;
SW3 = semidiurnal, westward propagating, zonal wavenumber 3 tide.

As discussed in Limpasuvan et al. (2016) and Stray, Orsolini, et al. (2015), the SSW onset criteria used here differ from the more widely used World Meteorological Organization definition based on 10-hPa zonal wind reversals at 60°N. However, Butler et al. (2015) discuss the range of previously used SSW definitions in the literature and emphasize the variability that different definitions can create in the timing of SSW onsets. With this in mind, Figure 7 shows the amplitude of each of the SDT components over each of the identified 13 SSWs. As in Figure 6, the data are presented over a 5-week period beginning 1 week before the SSW onset (day 0) as defined above.

The characteristic response of the three tidal components during the 2013 SSW presented in Figure 6 is evident in only a few rows of Figure 7. Examination of the other time series aligned by SSW onset date reveals no obvious clear and consistent pattern of SDT amplification locked to the SSW onset date. Nevertheless, there is a tendency for the SW2 component to reduce in amplitude following the SSW onset and then return strongly in the subsequent weeks apparent in a number of the time series. However, as mentioned earlier, the definition of an SSW carries a degree of subjectivity, and the exact onset date depends entirely on the parameter(s) used to define the SSW (Butler et al., 2015). In addition, although grouped together in a 13-event composite, these SSWs have very different characteristics. For example, estimating the duration of the zonal-mean zonal wind reversal at 50 km (averaged over latitude band 71–90°N), there is a range in these 13 SSWs from (approximately) 5 to 25 days, that is, a factor of 5. The implication is that this variable period when the zonal-mean zonal winds are reversed can condition the time development of the GW drag, the tidal waveguide, and the anomalous MMC. Furthermore, the SW2 components are observed by SuperDARN at a very limited latitude and altitude range. These limitations may not capture the tidal responses to SSWs (if any) beyond this region.

The complexity of the situation is further compounded by the range of onset dates of the 13 SSW events that span from 10 December to 22 February, with the majority SSW episodes occurring in December and January when the SDT is rapidly changing as evident in Figure 5. To mitigate these effects somewhat, the 30-day mean smoothed climatology of the amplitude of the individual SDT components (presented in Figure 7 as gray lines on the time series) is subtracted from each of the 13 events to generate a deseasonalized tidal amplitude anomaly. These anomaly time series are then averaged with respect to each SSW onset date to generate the composite tidal response to the SSW. Data in the individual time series are weighted by the reciprocal of the square of the standard error in the averaging process. Figure 8 presents this composite anomaly of SDT amplitudes covering the time period between 7 days before and 28 days after the SSW onset.

These composites reveal the average anomaly amplitude modulation of each of the measured SDT components. The most striking feature is a significant amplitude reduction of the SW2 component (reaching a minimum 5–6 days after the SSW onset) followed by an enhancement that peaks around 13 days after the SSW onset. The difference between these two extremes is nearly 10 m/s in amplitude as measured by the SuperDARN radars. During the period immediately after the SSW onset, we see a diminished SW2 amplitude which is coincident with a slight shift in the mean phase of the tide with a time of maximum up to 2 hr earlier (not shown). The subsequent amplitude enhancement of the SW2 tide, which peaks around 13 days after the SSW onset date, is accompanied by a small increase in the measured amplitude of the SW1 and SW3 components of the SDT by around 2–3 m/s.

As the migrating SDT (generated in the low-latitude lower atmosphere) propagates into the high-latitude MLT, the underlying wind field will significantly modulate its amplitude observed in the high-latitude MLT. In this case, when the underlying winds are strongly westward, conditions become less favorable for the propagation of the SW2 tide from the stratosphere into the MLT (see, e.g., Figure 2a in Limpasuvan et al., 2016). Riggin et al. (2003) discuss the refraction of the SDT in the middle atmosphere and note that the SW2 tide is only weakly propagating at high latitudes. Hence, even small perturbations in the background wind can inhibit the vertical propagation of the tide. We note that the reduction in the amplitude of the SDT immediately after the SSW onset is not observed in the low-latitude observations of Sridharan et al., (2012). This suggests that this amplitude modulation of the tide occurs in the tidal waveguide between the low and high latitudes. As the westward winds diminish and the upper stratospheric vortex recovers (on average around 10 days after the SSW onset), the zonal wind changes in the middle atmosphere associated with the return of normal wintertime vortex conditions allow the tidal perturbations to propagate freely into the high-latitude MLT. Hence, between 12 and 17 days after the SSW onset, the enhancement in tropical upper stratospheric ozone and the favorable underlying wind conditions throughout the NH upper stratosphere and mesosphere conspire to amplify the SW2 observed in the MLT. This amplification in the SW2 component is also accompanied by a smaller, though statistically significant, increase in the amplitude of both of the nonmigrating components measured here. Around 17 days after the SSW onset, as dynamical conditions and the corresponding ozone excess in the equatorial stratosphere return to normal (see, e.g., Figure 12 in Limpasuvan et al., 2016), the SW2 tidal anomaly returns to 0.

It is important to note that the event-to-event variability in duration discussed above may act to condition the tidal response to the SSW in a more complex way than can be captured through a simple superposed epoch approach, for example, through modulation in time of ozone at low latitudes or of the tidal waveguide. We further note that the observed results capture the tidal behavior over a rather limited MLT region near 60°N and 95 km. A detailed modeling study to further understand these processes and their influence on the tidal variability is currently underway.

## 6. Summary

Leveraging the extensive longitudinal coverage of the HF radar chain which comprises part of the NH SuperDARN network, this paper demonstrates a methodology to explicitly extract key components of tidal perturbations from meteor winds, not possible with single-station or uncoordinated ground observations. We present a 20-year climatology of nonmigrating SW1, migrating SW2, and nonmigrating SW3 components near 60°N and at 95 km altitude (in the MLT region). The annual variation of the derived tides concurs with previous results. Namely, the MLT SDT amplitude maximizes in the autumn equinox with a secondary wintertime maximum. These maxima are dominated by the migrating SW2, with smaller contributions from SW1 and SW3 especially around the equinoxes when the nonmigrating components are of comparable amplitude to the SW2 component.

The longitudinally spread, contemporaneous, long-term observations available from the cross-calibrated chain of similar SuperDARN radars allows us to separate the tidal components at high temporal resolution to demonstrate the behavior of the tide in response to rapid changes induced by the SSW phenomenon, a topic important to the understanding of the coupling processes in the MLT and potential ionospheric variability. In particular, the SW2 amplitude tends to be absent or anomalously weak during the time period immediately following the SSW onset (when the stratosphere wind reverses and becomes westward). However, as the polar atmosphere recovers from SSW conditions, the SW2 amplitudes become anomalously large which may be attributable to (1) the return of wind conditions that allow for upward tidal propagation and (2) anomalous tidal generation due to the enhancement of tropical stratosphere ozone (driven by upwelling and cooling associated with the mean circulation during SSW).

**Acknowledgments**
The authors acknowledge the use of SuperDARN data. SuperDARN is a collection of radars funded by national scientific funding agencies of Australia, Canada, China, France, Italy, Japan, Norway, South Africa, the United Kingdom, and the United States. SuperDARN data are available from Virginia Tech at vt.superdarn.org. R. E. H., P. J. E., and Y. O. are partly supported by the Research Council of Norway/CoE under contract 223252/F50. In addition to the support of a Fulbright Fellowship, V. L. is supported in part by the Large-Scale Dynamics Program at the National Science Foundation (NSF) under awards AGS-1642232 and AGS-1624068 and the Kerns Palmetto Professorship supported by the Coastal Carolina University's Provost office.

## References

Andrioli, V. F., Fritts, D. C., Batista, P. P., Clemesha, B. R., & Janches, D. (2013). Diurnal variation in gravity wave activity at low and middle latitudes. *Annales Geophysicae*, 31(11), 2123–2135. https://doi.org/10.5194/angeo-31-2123-2013

Angelats i Coll, M., & Forbes, J. M. (2002). Nonlinear interactions in the upper atmosphere: The $S = 1$ and $S = 3$ nonmigrating semidiurnal tides. *Journal of Geophysical Research*, 107(A8), 1157. https://doi.org/10.1029/2001JA900179

Baumgaertner, A. J. G., Jarvis, M. J., McDonald, A. J., & Fraser, G. J. (2006). Observations of the wavenumber 1 and 2 components of the semi-diurnal tide over Antarctica. *Journal of Atmospheric and Solar: Terrestrial Physics*, 68(11), 1195–1214. https://doi.org/10.1016/j.jastp.2006.03.001

Becker, E. (2017). Mean-flow effects of thermal tides in the mesosphere and lower thermosphere. *Journal of the Atmospheric Sciences*, 74(6), 2043–2063. https://doi.org/10.1175/JAS-D-16-0194.1

Beldon, C. L., & Mitchell, N. J. (2010). Gravity wave–tidal interactions in the mesosphere and lower thermosphere over Rothera, Antarctica (68°S, 68°W). *Journal of Geophysical Research*, *115*, D18101. https://doi.org/10.1029/2009JD013617

Bhattacharya, Y., Shepherd, G. G., & Brown, S. (2004). Variability of atmospheric winds and waves in the Arctic polar mesosphere during a stratospheric sudden warming. *Geophysical Research Letters*, *31*, L23101. https://doi.org/10.1029/2004GL020389

Burrage, M. D., Wu, D. L., Skinner, W. R., Ortland, D. A., & Hays, P. B. (1995). Latitude and seasonal dependence of the semidiurnal tide observed by the high-resolution Doppler imager. *Journal of Geophysical Research*, *100*(D6), 11,313–11,321. https://doi.org/10.1029/95JD00696

Butler, A. H., Seidel, D. J., Hardiman, S. C., Butchart, N., Birner, T., & Match, A. (2015). Defining sudden stratospheric warmings. *Bulletin of the American Meteorological Society*, *96*(11), 1913–1928. https://doi.org/10.1175/BAMS-D-13-00173.1

Chapman, S., & Lindzen, R. S. (1970). Atmospheric tides, D. In *Reidel Press*. Dordrecht: Holland.

Chisham, G., & Freeman, M. P. (2013). A reassessment of SuperDARN meteor echoes from the upper mesosphere and lower thermosphere. *Journal of Atmospheric and Solar - Terrestrial Physics*, *102*, 207–221. https://doi.org/10.1016/j.jastp.2013.05.018

de Wit, R. J., Hibbins, R. E., Espy, P. J., & Hennum, E. A. (2015). Coupling in the middle atmosphere related to the 2013 major sudden stratospheric warming. *Annales Geophysicae*, *33*(3), 309–319. https://doi.org/10.5194/angeo-33-309-2015

Espy, P. J., Jones, G. O. L., Swenson, G. R., Tang, J., & Taylor, M. J. (2004). Tidal modulation of the gravity wave momentum flux in the Antarctic mesosphere. *Geophysical Research Letters*, *31*, L11111. https://doi.org/10.1029/2004GLO19624

Farley, D. T. (1972). Multi-pulse incoherent scatter correlation function measurements. *Radio Science*, *7*(6), 661–666. https://doi.org/10.1029/RS007i006p00661

Goncharenko, L., Coster, A., Plumb, R., & Domeisen, D. (2012). The potential role of stratospheric ozone in the stratosphere-ionosphere coupling during stratospheric warmings. *Geophysical Research Letters*, *39*, L08101. https://doi.org/10.1029/2012GL051261

Greenwald, R. A., Baker, K. B., Dudeney, J. R., Pinnock, M., Jones, T. B., Thomas, E. C., et al. (1995). DARN/SuperDARN: A global view of the dynamics of high-latitude convection. *Space Science Reviews*, *71*(1-4), 761–796. https://doi.org/10.1007/BF00751350

Hagan, M. E., & Forbes, J. M. (2003). Migrating and nonmigrating semidiurnal tides in the upper atmosphere excited by tropospheric latent heat release. *Journal of Geophysical Research*, *108*(A2), 1062. https://doi.org/10.1029/2002JA009466

Hall, G. E., MacDougall, J. W., Moorcroft, D. R., St.-Maurice, J.-P., Manson, A. H., & Meek, C. E. (1997). Super Dual Auroral Radar Network observations of meteor echoes. *Journal of Geophysical Research*, *102*(A7), 14,603–14,614. https://doi.org/10.1029/97JA00517

Hibbins, R. E., Freeman, M. P., Milan, S. E., & Ruohoniemi, J. M. (2011). Winds and tides in the mid-latitude Southern Hemisphere upper mesosphere recorded with the Falkland Islands SuperDARN radar. *Annales Geophysicae*, *29*(11), 1985–1996. https://doi.org/10.5194/angeo-29-1985-2011

Hibbins, R. E., & Jarvis, M. J. (2008). A long-term comparison of wind and tide measurements in the upper mesosphere recorded with an imaging Doppler interferometer and SuperDARN radar at Halley, Antarctica. *Atmospheric Chemistry and Physics*, *8*(5), 1367–1376. https://doi.org/10.5194/acp-8-1367-2008

Hibbins, R. E., Marsh, O. J., McDonald, A. J., & Jarvis, M. J. (2010). A new perspective on the longitudinal variability of the semidiurnal tide. *Geophysical Research Letters*, *37*, L14804. https://doi.org/10.1029/2010GL044015

Hussey, G. C., Meek, C. E., Andre, D., Manson, A. H., Sofko, G. J., & Hall, C. M. (2000). A comparison of Northern Hemisphere winds using SuperDARN meteor trail and MF radar wind measurements. *Journal of Geophysical Research*, *105*(D14), 18,053–18,066. https://doi.org/10.1029/2000JD900272

Iimura, H., Fritts, D. C., Wu, Q., Skinner, W. R., & Palo, S. E. (2010). Nonmigrating semidiurnal tide over the Arctic determined from TIMED Doppler Interferometer wind observations. *Journal of Geophysical Research*, *115*, D06109. https://doi.org/10.1029/2009JD012733

Iimura, H., Palo, S. E., Wu, Q., Killeen, T. L., Solomon, S. C., & Skinner, W. R. (2009). Structure of the nonmigrating semidiurnal tide above Antarctica observed from the TIMED Doppler Interferometer. *Journal of Geophysical Research*, *114*, D11102. https://doi.org/10.1029/2008JD010608

Kishore Kumar, G., & Hocking, W. K. (2010). Climatology of northern polar latitude MLT dynamics: Mean winds and tides. *Annales Geophysicae*, *28*(10), 1859–1876. https://doi.org/10.5194/angeo-28-1859-2010

Kleinknecht, N. H., Espy, P. J., & Hibbins, R. E. (2014). Climatology of zonal wave numbers 1 and 2 planetary wave structure in the MLT using a chain of Northern Hemisphere SuperDARN radars. *Journal of Geophysical Research: Atmospheres*, *119*, 1292–1307. https://doi.org/10.1002/2013JD019850

Limpasuvan, V., Orsolini, Y. J., Chandran, A., Garcia, R. R., & Smith, A. K. (2016). On the composite response of the MLT to major sudden stratospheric warming events with elevated stratopause. *Journal of Geophysical Research: Atmospheres*, *121*, 4518–4537. https://doi.org/10.1002/2015JD024401

Lin, J. T., Lin, C. H., Chang, L. C., Huang, H. H., Liu, J. Y., Chen, A. B., et al. (2012). Observational evidence of ionospheric migrating tide modification during the 2009 stratospheric sudden warming. *Geophysical Research Letters*, *39*, L02101. https://doi.org/10.1029/2011GL050248

Manson, A. H., Meek, C. E., Chshyolkova, T., Xu, X., Aso, T., Drummond, J. R., et al. (2009). Arctic tidal characteristics at Eureka (80°N, 86°W) and Svalbard (78°N, 16°E) for 2006/07: Seasonal and longitudinal variations, migrating and non-migrating tides. *Annales Geophysicae*, *27*(3), 1153–1173. https://doi.org/10.5194/angeo-27-1153-2009

Manson, A. H., Meek, C. E., Xu, X., Aso, T., Drummond, J. R., Hall, C. M., et al. (2011). Characteristics of Arctic tides at CANDAC-PEARL (80°N, 86°W) and Svalbard (78°N, 16°E) for 2006–2009: Radar observations and comparisons with the model CMAM-DAS. *Annales Geophysicae*, *29*, 1939–1954. https://doi.org/10.5194/angeo-29-1939-2011

Marsh, D. R. (2011). Chemical–dynamical coupling in the mesosphere and lower thermosphere. In *Aeronomy of the Earth's Atmosphere and Ionosphere*, IAGA Spec. Sopron Book Ser., (Vol. 2, pp. 3–17). Netherlands: Springer.

Marsh, D. R., Mills, M. E., Kinnison, D. E., Lamarque, J.-F., Calvo, N., & Polvani, L. M. (2013). Climate change from 1850 to 2005 simulated in CESM1 (WACCM). *Journal of Climate*, *26*(19), 7372–7391. https://doi.org/10.1175/JCLI-D-12-0558.1

Mitchell, N. J., Pancheva, D., Middleton, H. R., & Hagan, M. E. (2002). Mean winds and tides in the Arctic mesosphere and lower thermosphere. *Journal of Geophysical Research*, *107*(A1), 1004. https://doi.org/10.1029/2001JA900127

Miyahara, S., Yoshida, Y., & Miyoshi, Y. (1993). Dynamic coupling between the lower and upper atmosphere by tides and gravity waves. *Journal of Atmospheric and Terrestrial Physics*, *55*(7), 1039–1053. https://doi.org/10.1016/0021-9169(93)90096-H

Miyoshi, Y., Pancheva, D., Mukhtarov, P., Jin, H., Fujiwara, H., & Shinagawa, H. (2017). Excitation mechanism of non-migrating tides. *Journal of Atmospheric and Solar: Terrestrial Physics*, *156*, 24–36. https://doi.org/10.1016/j.jastp.2017.02.012

Murphy, D. J., Forbes, J. M., Walterscheid, R. L., Hagan, M. E., Avery, S. K., Aso, T., et al. (2006). A climatology of tides in the Antarctic mesosphere and lower thermosphere. *Journal of Geophysical Research*, *111*, D23104. https://doi.org/10.1029/2005JD006803

Nozawa, S., Hall, C. M., Tsutsumi, M., Brekke, A., Ogawa, Y., Tsuda, T. T., et al. (2012). Mean winds, tides, and quasi-2 day waves above Bear Island. *Journal of Atmospheric and Solar: Terrestrial Physics*, *90-91*, 26–44. https://doi.org/10.1016/j.jastp.2012.05.002

Orsolini, Y. J., Limpasuvan, V., Pérot, K., Espy, P. J., Hibbins, R. E., Lossow, S., et al. (2017). Modelling the descent of nitric oxide during the elevated stratopause event of January 2013. *Journal of Atmospheric and Solar: Terrestrial Physics*, *155*, 50–61. https://doi.org/10.1016/j.jastp.2017.01.006

Pedatella, N. M., & Forbes, J. M. (2010). Evidence for stratosphere sudden warming-ionosphere coupling due to vertically propagating tides. *Geophysical Research Letters*, *37*, L11104. https://doi.org/10.1029/2010GL043560

Portnyagin, Y. I., Solovjova, T. V., Makarov, N. A., Merzlyakov, E. G., Manson, A. H., Meek, C. E., et al. (2004). Monthly mean climatology of the prevailing winds and tides in the Arctic mesosphere/lower thermosphere. *Annales Geophysicae*, *22*(10), 3395–3410. https://doi.org/10.5194/angeo-22-3395-2004

Press, W. H., Flannery, B. P., Teukolsky, S. A., & Vetterling, W. T. (1992). *Numerical recipes in C: The art of scientific computing* (2nd ed.). Cambridge University Press.

Randel, W. J. (1993). Global variations of zonal mean zonal ozone during stratospheric warming events. *Journal of the Atmospheric Sciences*, *50*(19), 3308–3321. https://doi.org/10.1175/1520-0469(1993)050<3308:GVOZMO>2.0.CO;2

Riggin, D. M., Meyer, C. K., Fritts, D. C., Jarvis, M. J., Murayama, Y., Singer, W., et al. (2003). MF radar observations of seasonal variability of semidiurnal motions in the mesophere at high northern and southern latitudes. *Journal of Atmospheric and Solar: Terrestrial Physics*, *65*(4), 483–493. https://doi.org/10.1016/S1364-6826(02)00340-1

Sassi, F., Liu, H.-L., Ma, J., & Garcia, R. R. (2013). The lower thermosphere during the Northern Hemisphere winter of 2009: A modeling study using high-altitude data assimilation products in WACCM-X. *Journal of Geophysical Research: Atmospheres*, *118*, 8954–8968. https://doi.org/10.1002/jgrd.50632

Smith, A. K. (2012). Global dynamics of the MLT. *Surveys in Geophysics*, *33*(6), 1177–1230. https://doi.org/10.1007/s10712-012-9196-9

Sridharan, S., Sathishkumar, S., & Gurubaran, S. (2012). Variabilities of mesospheric tides during sudden stratospheric warming events of 2006 and 2009 and their relationship with ozone and water vapour. *Journal of Atmospheric and Solar: Terrestrial Physics*, *78-79*, 108–115. https://doi.org/10.1016/j.jastp.2011.03.013

Stray, N. H., de Wit, R. J., Espy, P. J., & Hibbins, R. E. (2014). Observational evidence for temporary planetary-wave forcing of the MLT during fall equinox. *Geophysical Research Letters*, *41*, 6281–6288. https://doi.org/10.1002/2014GL061119

Stray, N. H., Espy, P. J., Limpasuvan, V., & Hibbins, R. E. (2015). Characterisation of quasi-stationary planetary wave in the MLT during summer. *Journal of Atmospheric and Solar: Terrestrial Physics*, *127*, 30–36. https://doi.org/10.1016/j.jastp.2014.12.003

Stray, N. H., Orsolini, Y. J., Espy, P. J., Limpasuvan, V., & Hibbins, R. E. (2015). Observations of planetary waves in the mesosphere-lower thermosphere during stratospheric warming events. *Atmospheric Chemistry and Physics*, *15*(9), 4997–5005. https://doi.org/10.5194/acp-15-4997-2015

Teitelbaum, H., & Vial, F. (1991). On tidal variability by nonlinear interaction with planetary waves. *Journal of Geophysical Research*, *96*(A8), 14,169–14,178. https://doi.org/10.1029/91JA01019

Tweedy, O. V., Limpasuvan, V., Orsolini, Y. J., Smith, A. K., Garcia, R. R., Kinnison, D., et al. (2013). Nighttime secondary ozone layer during major stratospheric sudden warmings in specified-dynamics WACCM. *Journal of Geophysical Research: Atmospheres*, *118*, 8346–8358. https://doi.org/10.1002/jgrd.50651

Ward, W. E. (1998). Tidal mechanisms of dynamical influence on oxygen re-combination airglow in the mesosphere and lower thermosphere. *Advances in Space Research*, *21*(6), 795–805. https://doi.org/10.1016/S0273-1177(97)00676-5